\documentclass[rmp,twocolumn,floats]{revtex4}

\usepackage{graphics,graphicx,epsfig}
\usepackage{amssymb}
\usepackage{epsf,epstopdf,wrapfig}
\usepackage{color}
\usepackage{natbib} 

\usepackage{natbib}

\newcommand{\beq}{\begin{equation}}
\newcommand{\eeq}{\end{equation}}
\newcommand{\beqn}{\begin{eqnarray}}
\newcommand{\eeqn}{\end{eqnarray}}
\newcommand {\e}[1]{\mathrm{~#1}}

\begin{document}

\title{Retinal adaptation and invariance to changes in higher-order stimulus statistics}

\author{Ga\v{s}per Tka\v{c}ik\footnote{gtkacik@ist.ac.at}$^{a}$, Anandamohan Ghosh$^{b}$, Elad Schneidman$^c$, and Ronen Segev$^{d}$}

\affiliation{$^a$Institute of Science and Technology Austria, Am Campus 1, A-3400 Klosterneuburg, Austria\\
$^b$Indian Institute of Science Education and Research-Kolkata, Mohanpur, 741252 Nadia, India\\
$^c$Department of Neurobiology, Weizmann Institute of Science, 76100 Rehovot, Israel\\
$^d$Facutly of Natural Sciences, Department of Life Sciences and Zlotowski Center for Neuroscience, Ben Gurion University of the Negev, 84105 Be'er Sheva, Israel}

\date{\today}

\begin{abstract}
Adaptation in the retina is thought to optimize the encoding of natural light signals into sequences of spikes sent to the brain. However, adaptation also entails computational costs: adaptive code is intrinsically ambiguous, because output symbols cannot be trivially mapped back to the stimuli without the knowledge of the adaptive state of the encoding neuron. It is thus important to learn which statistical changes in the input do, and which do not, invoke adaptive responses, and ask about the reasons for potential limits to adaptation. We measured the ganglion cell responses in the tiger salamander retina to controlled changes in the second (contrast), third (skew) and fourth (kurtosis) moments of the light intensity distribution of spatially uniform  temporally independent stimuli. The skew and kurtosis of the stimuli were chosen to cover the range observed in natural scenes. We quantified adaptation in  ganglion cells by studying two-dimensional linear-nonlinear models that  capture well the retinal encoding properties across all stimuli. We found that the retinal ganglion cells adapt to contrast, but exhibit remarkably invariant behavior to changes in higher-order statistics. Finally, by theoretically analyzing optimal coding in LN-type models, we showed that the neural code can maintain a high information rate  without dynamic adaptation despite changes in stimulus skew and kurtosis.
\end{abstract}

\maketitle

\section{Introduction}

Adaptation is ubiquitous in the nervous system, from synaptic depression \citep{tsodyks97,abbott97} and single neuron spiking \citep{adrian28,partridge76}, to the activity of neural modules (e.g. \citet{muller99}). In sensory systems, it has been suggested to be a key design principle of the neural code \citep{wark07}, which may allow for optimal information coding by matching the neural responses to stimulus statistics \citep{attneave54,barlow61,atick90,atick92}. The retina is one of the most studied highly adaptive neural circuits, in which the mapping between stimuli and neural response  changes to match the statistics of the mean light intensity \citep{shapley84}, temporal and spatial contrast and spatial scale \citep{beaudoin07,chander01,smirnakis97}, pattern \citep{hosoya05}, relative motion \citep{olveczky07} and periodicity \citep{schwartz08}. 

Since adaptation requires some form of memory and inference of the stimulus statistics to which the system should adapt, the mechanism and nature of adaptation have been studied extensively. For example, the dynamic structure of the retinal ganglion cell receptive fields \citep{srinivasan82}, and contrast adaptation in the vertebrate and fly visual systems \citep{victor86,laughlin81,smirnakis97,brenner00,chander01,shapley79} have been characterized as gain-control mechanisms that serve to efficiently encode the variation of the stimulus around the mean into a limited dynamic range of firing rates at the output. It has been further shown that neural systems adapt not only to various stationary stimuli, but also to dynamic changes in stimulus distributions taking place across multiple timescales \citep{ulanovsky04,wallach08,fairhall01}.

Despite its ubiquitous presence, it is still not clear what are the limits to adaptation, and in particular, which stimulus changes should lead to adaptive responses and which should not. Moreover, by its nature, adaptation comes with an inherent caveat or cost \citep{fairhall01}: adaptive code is ambiguous, which requires the decoder to keep some knowledge of the coding context. Since most studies of adaptation analyzed neural systems' response to first- and second-order spatio-temporal statistics in the stimulus, we addressed here the nature of neural response to changes in higher-order structure of visual stimuli; such higher-order structure is characteristic of natural scenes \citep{simoncelli01,geisler08} and is perceptually salient \citep{portilla00,chubb04,tkacik11}.

To characterize adaptation and invariance to stimulus statistics beyond luminance and contrast, we studied retinal responses to spatially uniform stimuli where light intensities were drawn independently from distributions with tunable amounts of  skewness and kurtosis. We analyzed salamander retinal ganglion cell responses using maximally informative dimensions, and quantified the encoding properties of the neurons using 2D linear-nonlinear (LN) models. 
We found clear signatures of contrast adaptation in the rescaling of nonlinearities to these stimuli, but also that the same neurons display striking encoding invariance to changes in stimulus skewness or kurtosis, even when they are exposed to strongly bimodal stimulus distributions.

\section{Materials and methods}
{\footnotesize

{\bf Natural image statistics.} To sample the range of naturally occurring values for contrast, skewness and kurtosis, we took a sample of 501 calibrated grayscale images from the Penn Natural Image Database (PNIDb) \citep{tkacik11}. From each image we selected 400 random patches $200 \times 200$ pixels in size, which corresponds in area to the angular size of about 3 degrees, roughly the size of the center receptive field of a salamander retinal ganglion cell. Averaging over each patch to get the mean luminance in that patch, we computed the contrast, skewness and kurtosis of the distribution of patch luminances in a given image. Repeating the process over all images in our selection (containing shots of Baboon habitat in Okavango delta in Botswana, including landscape images, some with horizon, closeups of the ground, and a small selection of man-made objects in that habitat), we accumulated  natural distributions of contrast, skewness and kurtosis.

In our analyses, contrast is defined as $C=\sigma_L/\bar{L}$, where $\bar{L}$ is the mean luminance, $\sigma_L=\sqrt{\langle (L-\bar{L})^2\rangle}$ is the std of the luminance distribution $P(L)$ and brackets denote averaging over this distribution; skewness $S=\langle (L-\bar{L})^3 \rangle / \sigma_L^3$; and kurtosis $K=\langle (L-\bar{L})^4 \rangle / \sigma_L^4-3$. Note that kurtosis is defined to be 0 for  Gaussian distributions.
\\

{\bf Electrophysiology.} Multi-electrode array recordings were performed on adult tiger salamander (\emph{Ambystoma tigrinum}) \citep{meister94}. All experiments were done according the regulation of Ben Gurion University of the Negev and the laws of the State of Israel. Prior to the experiment the salamander was adapted to bright light for 30 minutes. Retinas were isolated from the eye and peeled from the sclera together with the pigment epithelium. Retinas were placed with the ganglion cell layer facing a multielectrode array with 252 electrodes (Ayanda Biosystems, Switzerland) and superfused with oxygenated (95\% O$_2$, 5\% CO$_2$) Ringer medium which contained 110mM NaCl, 22mM NaHCO$_3$, 2.5mM KCl, 1mM CaCl$_2$, 1.6mM MgCl$_2$, and 18mM glucose, at room temperature. The electrode diameter was 10$\mathrm{\mu m}$ and electrode spacing varied between 40 and 80 $\mathrm{\mu m}$. Recordings of 24-30 hours were achieved consistently. Extracellularly recorded signals were amplified (MultiChannel Systems, Germany), digitized at 10 kHz on four personal computers and stored for off-line spike sorting and analysis. Spike sorting was done by extracting from each potential waveform the amplitude and width, followed by manual clustering using an in-house program written in MATLAB.
\\

{\bf Stimulation.} The stimulus was projected onto the salamander retina from a CRT video monitor (ViewSonic G90fB) at a frame rate of 60Hz such that each stimulus frame was presented twice in a row (for a stimulus sampling rate of 30Hz) using standard optics. The stimulus intensity was presented in grayscale and was gamma corrected for the monitor. Gaussian stimulus distributions with the desired variance were generated using MATLAB random number generator. We refer to all non-Gaussian stimuli as HOS (higher-order statistics) stimuli, which we generated with the statistics given in Table~\ref{t1} as follows. The kurtotic distributions were sums of two equal-variance Gaussian distributions of equal weight and  displaced means. The skewed stimuli are sums of two Gaussian distributions with unequal variances. All resulting distributions have the same mean luminance of $\bar{L}\approx 225 \e{lux}$. 

We performed two experiments. In the first (23 cells), all 9 stimuli were displayed in long, non-repeated sequences (52202 frames of $33.33\e{ms}$ each for each of the 9 stimuli), allowing us to infer LN models precisely; we used the Gaussian stimulus to fit LN models using both the spike-triggered average / covariance and maximally informative dimensions, to check how closely the two inference methods agree. In the second experiment (40 cells), the two Gaussian stimuli were absent, while for the 7 remaining HOS stimuli  each non-repeated sequence was followed by a repeated sequence (30 repeats, 602 frames at $33.33\e{ms}$ for each repeat), used to validate our models.
\begin{table}[tbh]
\vspace{0.5cm}
\begin{tabular}{|l|c|c|c|c|}
\hline
stimulus $\mathcal{S}$ & symbol	&	contrast  $C$	&	 skewness $S$ 	&	 kurtosis $K$\\ \hline\hline
{\bf Gaussian} &    C+ &   {\bf 0.097}        &	0	                   & 0 \\ \hline
		&    C++ &  {\bf 0.177}         & 	0			& 0 \\ \hline \hline
{\bf Skewed}    
		& S{-}{-} &	0.170	&	{\bf -1.9}			&	5.1	\\ \hline 
(HOS)		&    S- & 0.172		&	{\bf -1.0}			&	6.0	\\ \hline
		&     S+ &0.175		&	{\bf 1.0}			&      5.2 \\ \hline
		&   S++ & 0.178		& 	{\bf 1.9}			&     5.2 \\ \hline \hline

{\bf Kurtotic}    &    K{-}{-} & 0.177		&	0			&	{\bf -1.8}	\\ \hline
	(HOS)	&   K- &0.176		&	0			&	{\bf -0.9}	\\\hline
	&	K+ &0.173	&	0			&	{\bf 5.3} \\ \hline
\end{tabular}
\caption{{\bf Stimuli $\mathcal{S}$ used in the experiment (see main text for the definition of the statistics $C, S, K$)}. The shorthand symbol for the stimulus starts with the C/S/K (for contrast, skew, kurtosis) and is followed by {-},{-}{-},+,++ (small magnitude and negative, large magnitude and negative, small magnitude and positive, large magnitude and positive); therefore, $\mathcal{S}=\{$C+,C++,S{-}{-},S-,S+,S++,K{-}{-},K-,K+$\}$. Parameters in the table denoted in bold were varied in each of the three stimulus categories.}
\label{t1}
\end{table}
\\

{\bf Linear filters.} In inferring LN encoding models, reverse correlation techniques cannot directly be applied to non-Gaussian stimuli because they lead to biased filter estimates. Instead, we used maximally informative dimensions (MID) \citep{sharpee04}. To look for a single significant filter $\mathbf{k}_1$, one performs the following maximization over possible linear filters $\mathbf{k}_1$, constrained to unit norm ($\mathbf{k}_1 \cdot \mathbf{k}_1=1$):
\begin{equation}
I_{\mathrm{spike}} = \mathrm{max}_{\mathbf{k}_1}\;D_{KL}\left( P( \mathbf{k}_1 \cdot \mathbf{s} | \mathrm{spike}) || P( \mathbf{k}_1 \cdot \mathbf{s})\right). \label{eqinfo}
\end{equation}
Here $D_{KL}$ is the Kullback-Leibler divergence \citep{shannon} between the spike-triggered distribution  and the prior distribution of stimulus fragment projections onto $\mathbf{k}_1$, and $\mathbf{s}$ are stimulus fragments (those preceding the spike for the spike-triggered distribution, and all fragments for the prior distribution). To look for 2D models, we repeated the same optimization with two filters $\{\mathbf{\tilde{k}}_1, \mathbf{\tilde{k}}_2\}$; the spike-triggered and prior distributions are two-dimensional in this case. For all neurons, the single most informative filter $\mathbf{k}_1$ was contained in the space of the 2 filters $\{\mathbf{\tilde{k}}_1, \mathbf{\tilde{k}}_2\}$; for further analysis, we rotated the system of reference such that the first filter was the single most informative filter $\mathbf{k}_1$ (which mostly corresponded to the spike-triggered average for those cells that were exposed to Gaussian stimulus), while the second filter $\mathbf{k}_2$ spanned the  $\{\mathbf{k}_1, \mathbf{\tilde{k}}_2\}$ subspace together with $\mathbf{k}_1$, and formed an orthonormal basis, $\mathbf{k}_1\cdot\mathbf{k}_2=0$, $\mathbf{k}_2\cdot\mathbf{k}_2=1$. 

The filters extended over $600\e{ms}$ and were sampled on 36 equidistant points, with temporal resolution of $16.67\e{ms}$. We expressed the filters as a combination of 16 basis functions, $\mathbf{k}_{1,2}=\sum_{\mu=1}^{16} \alpha^{\mu}_{1,2} \mathbf{b}_{\mu}$, where $\mathbf{b}_{\mu}$ are unit-area Gaussian bumps with $40\e{ms}$ width, uniformly tiling the $600\e{ms}$ span of the filters, and $\alpha$ are the expansion coefficients; we maximized $I_{\mathrm{spike}}$ in the space of parameters $\alpha$. This expansion made the filters smooth and very slightly improved generalization performance, but the results were stable even if we inferred directly in the space of $\mathbf{k}$. 

For performance reasons we estimated $D_{KL}$ during MID optimization runs using kernel-smoothing estimation; for final results we used the  context-weighted-tree (CTW) estimator \citep{sadeghi09}; the two estimators agreed without bias and to within $4\%$  for final filters across all cells and stimuli. Optimization was done using custom stochastic gradient descent code that can avoid local maxima. We performed two optimization runs for each cell and each stimulus, and the values of information per spike between the two runs differed by 1\% on average, $98.6\%$ of the runs had a difference smaller than $5\%$. 

To quantitatively compare the shapes of the filters across stimuli in experiment 1, we needed to ensure that each stimulus condition had enough spikes for good filter inference. We required each cell to have an average firing rate of at least $1.5\e{Hz}$; 15 out of 23 cells passed this cut. In experiment 1 we displayed Gaussian stimuli in addition to HOS stimuli, and we computed spike-triggered average / covariance to extract STA and the next most significant filter (orthogonal to the STA) from Gaussian segments. To judge the significance of the eigenvectors in the STC analysis we used bootstrapping with subsets of recorded spikes following \citet{rdr}.
\\

{\bf Nonlinearities and PSTH prediction.} After having reconstructed the linear filters $\{\mathbf{k}_1,\mathbf{k}_2\}$, we estimated the nonlinearities as follows: $\mathcal{N}(v_1,v_2)= \bar{r}\;P(v_1,v_2|\mathrm{spike}) / P(v_1,v_2) $, where $v_{1,2}=\mathbf{k}_{1,2} \cdot \mathbf{s}$ are the projections of the stimulus onto the two linear filters, and $\bar{r}$ is the mean firing rate of the neuron in a given stimulus condition. For 2D nonlinearities, we binned $\{v_1,v_2\}$ values on a $16 \times 16$ grid; for estimating 1D projections of the full 2D nonlinearity, we binned into a number of bins that was adaptively dependent on the number of spikes, and used kernel smoothing to approximate the probability distributions. Prediction performance was only slightly changed when 2D nonlinearities were sampled over $32 \times 32$ domain, and in general dropped due to overfitting when $64\times 64$ bins were used. We used  2D LN models fit on the nonrepeated segment of the stimulus to predict the PSTH for the repeated segments, using the time resolution of $16.67\e{ms}$, half the stimulus refresh rate. The fit was quantified by computing the Pearson cross-correlation between the true and predicted PSTH.
\\

{\bf Information captured by the models.} In the framework of LN encoding models, one assumes that only a small number $K$ of linear projections $\{v_1,v_2,\dots,v_K\}$ of a high-dimensional stimulus $\mathbf{s}$ determine whether a neuron spikes or not \citep{baa1}. In other words, the neuron is viewed as implementing a probabilistic dependency chain: $\mathbf{s}\rightarrow\{v_1,v_2,\dots,v_K\}\rightarrow \mathcal{N}(v_1,v_2,\dots,v_K)\rightarrow\mathrm{spike}$, which implies a chain of information processing inequalities: $I_{\mathrm{spike}}(\mathbf{s};\mathrm{spike}) \geq  I_{\mathrm{spike}}(\{v_1,v_2,\dots,v_K\};\mathrm{spike})  \geq  I_{\mathrm{spike}}(\mathcal{N}(v_1,v_2,\dots,v_K);\mathrm{spike})$. It is possible to estimate $I_{\mathrm{spike}}(\mathbf{s};\mathrm{spike})$ from repeated presentations of the same stimulus. If $r(t)=\langle \sum_{\mu=1}^{N(\rho)} \delta(t-t^\rho_\mu)\rangle_{\rm repeats}$ is the time-dependent firing rate, where $\rho=1,\dots,R$ indexes the repeats, $t^\rho_\mu$ is the time of $\mu$-th spike in repeat $\rho$, $N(\rho)$ is the total number of spikes in repeat $\rho$, and $t\in [0,T]$ denotes time within the repeat of length $T$, the estimate for true information per spike is given by \citep{brenner00}: 
\begin{equation}
I_{\rm spike}^{ub} = \frac{1}{T}\int_0^T dt\; \frac{r(t)}{\bar{r}}\log_2\frac{r(t)}{\bar{r}}; \label{brenner}
\end{equation}
here $\bar{r}=1/T \int_0^T dt\; r(t)$ is the average firing rate across the repeated stimulus segment. This quantity is an upper bound to the information quantities defined above for LN models.  The fraction, e.g. $I_{\rm spike}(\{v_1,v_2\} | {\rm spike})/I_{\rm spike}^{ub}$ (between 0 and 1), tells us how well the two stimulus projections capture the full dependence of spiking on the stimulus. Similarly, $I_{\rm spike}(\mathcal{N}(v_1,v_2) | {\rm spike})/I_{\rm spike}^{ub}$ (which needs to be lower or equal to the information in the two projections for the same neuron and stimulus) quantifies how much information is further lost when compressing the description of spike-dependence from two projections into a single nonlinear combination. 

The  information was estimated from Eq~(\ref{brenner}) with rates computed in $16.67\e{ms}$ bins (matching the time resolution of the temporal filters and the sampling used to calculate $D_{KL}$ in Eq~(\ref{eqinfo})), and was corrected for small-sample bias by repeatedly estimating the information on random subsets of stimulus repeats of varying sizes, plotting the information estimates against $1/N_{\rm repeats}$ and extrapolating to infinite number of repeats; we also applied the correction for the difference between mean firing rates in the repeated and non-repeated stimulus segments \citep{fairhall06}. The expected extrapolation error is below $1\%$. To obtain an upper bound for the systematic error due to short repeat length, we split the repeated segment in half and estimated the information separately on each half, which resulted in  relative differences with a std of $8\%$; we expect the true error to be smaller. Information rates were estimated by computing the information per spike and multiplying by the mean firing rate.

For all neurons recorded in experiment 2 (with non-repeated and repeated stimuli), we computed several information-theoretic quantities: (i+ii) $I_{\rm spike}(\{v_1,v_2\} | {\rm spike}, \mathcal{S})/I_{\rm spike}^{ub}(\mathcal{S})$ is the information fraction captured by 2 (and 1, respectively) linear filter(s), fit \emph{separately} to each stimulus condition $\mathcal{S}$;  (iii+iv) $I_{\rm spike}(\mathcal{N}(v_1,v_2) | {\rm spike}, \mathcal{S})/I_{\rm spike}^{ub}(\mathcal{S})$ are the fractions captured by the nonlinear combination of 2 (and 1, respectively) projection(s) fit \emph{separately} to each stimulus condition $\mathcal{S}$; (v+vi) $I_{\rm spike}(\{v_1,v_2\} | {\rm spike}, \mathrm{global})/I_{\rm spike}^{ub}(\mathcal{S})$ and $I_{\rm spike}(\mathcal{N}(v_1,v_2) | {\rm spike}, \mathrm{global})/I_{\rm spike}^{ub}(\mathcal{S})$ are the fractions captured by a single 2D model (by two projections and their nonlinear combination, respectively) that has been fit \emph{globally} to all stimulus stimuli $\mathcal{S}$. These quantities were all estimated using CTW estimator for Kullback-Leibler divergence. When estimated on spike trains that have been shuffled with respect to the stimulus, the estimator yields negligible values below $10^{-3}$ bits.
\\

{\bf Predicted information rate of LN models.} To computationally simulate the effect that the (lack of) adaptation would have on the information rate when the stimulus statistics changes, we used the LN models inferred at high contrast C++ to  predict the firing rate $r(t)$ for stimuli with contrasts C$\,<\,$C++ and ask how much information such neurons would carry per spike in the absence of any adaptation. This information in the predicted rate, $I_{\rm rate}$, was evaluated using Eq~(\ref{brenner}) and expressed as a fraction of information captured by the two relevant filters (which does not depend on contrast and only serves as a normalization). We similarly asked how the same quantity would behave when the global models (single LN models for all cells that are fit across all stimuli, and therefore have no adaptation) were used to encode information into the rate for each of the skewed stimuli.
}

\section{Results}
\begin{figure*}[tbh] 
\centering
\vspace{0.5cm}
\includegraphics[width=5in]{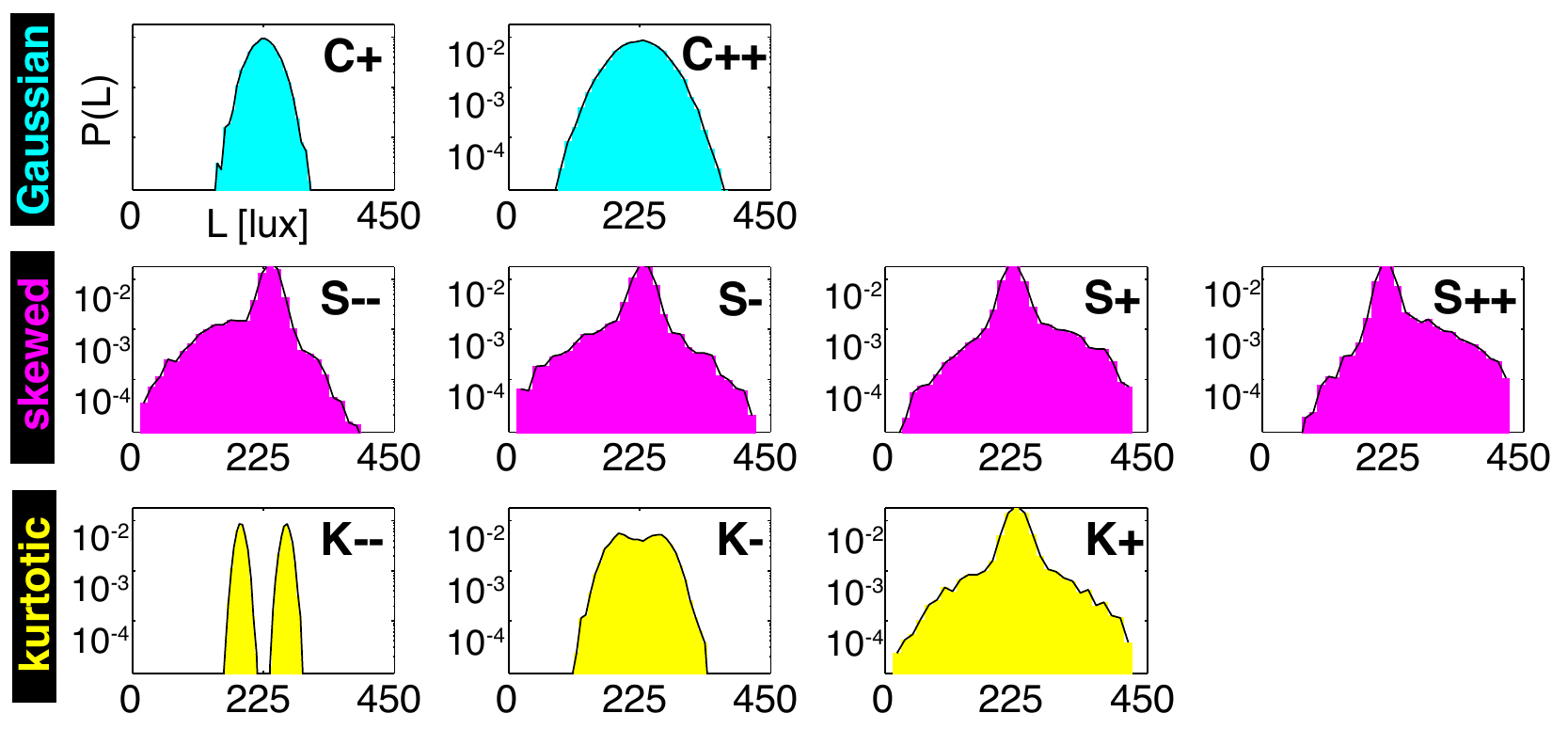} 
\caption{{\bf Synthetic stimuli used to probe salamander retinal ganglion cells.} The probability densities, $\log P_\mathcal{S}(L)$, for all 9 stimuli $\mathcal{S}$ used, grouped into 3 categories (cyan = Gaussian, magenta = skewed, and yellow = kurtotic). All stimuli are matched in mean (225 lux), and all except for C+ have the same contrast; for details, see Table~\ref{t1}. } 
\label{f1}
\end{figure*}

\begin{figure*}[tbh] 
\centering
\vspace{0.5cm}
\includegraphics[width=5in]{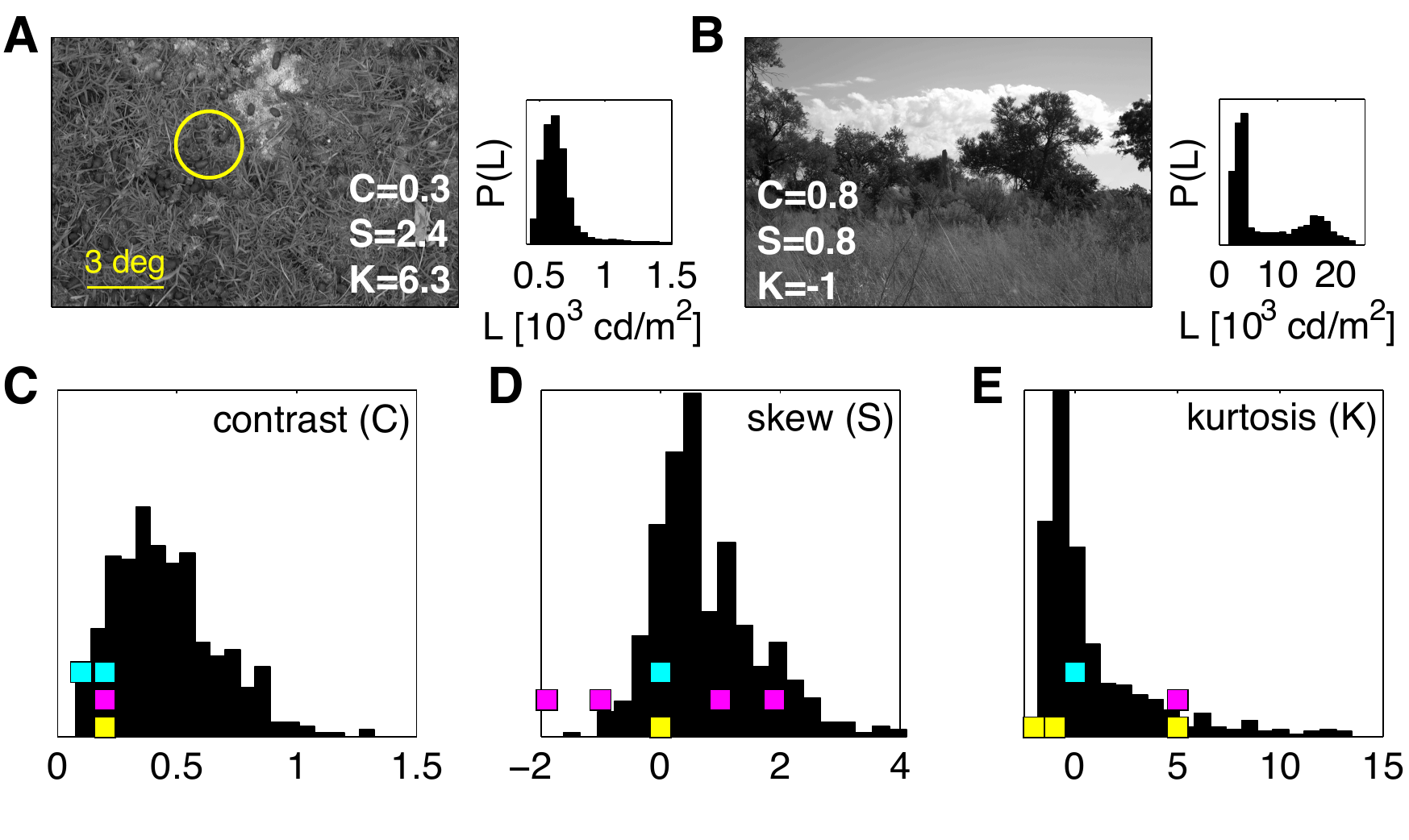} 
\caption{{\bf Higher-order statistics in natural scenes.} {\bf A,B)} Two example images from the Penn Natural Image Database. The grayscale images are calibrated into units of cd/m$^2$. The yellow circle represents the typical size of the salamander retinal ganglion cell center. Luminance was averaged in patches of this size, and contrast ($C$), skewness ($S$) and kurtosis ($K$) were computed for the distribution over many patches from each image. The distributions $P(L)$ for the two example images are  shown as insets, and the corresponding values for $C,S,K$ are displayed in the two image panels. {\bf C,D,E)} The distribution of contrast, skewness and kurtosis, respectively, over 501 natural images. Colored squares represent the values of the three parameters used in synthetic stimuli (color coded as in Fig~\ref{f1}). 2 cyan stimuli differ in contrast $C$ but have constant $S$ and $K$; 4 magenta stimuli differ in skew $S$ but have constant values of $C$ and $K$; and 3 yellow stimuli differ in kurtosis $K$ but have constant $C$ and $S$ (see Table~\ref{t1}).} 
\label{f2}
\end{figure*}

To characterize how the retina encodes higher-order statistics (HOS) of the luminance distribution, we presented it with a set of 9 synthetic spatially homogenous stimuli $\mathcal{S}$, where the light intensity of each stimulus frame was drawn independently from distributions $P_{\mathcal{S}}(L)$ that were matched in mean $\bar{L}$ (see Materials and Methods). The stimuli  differed systematically in contrast, skewness, and kurtosis, as depicted in Fig~\ref{f1}. The particular values were chosen  to span the  range of skewness and kurtosis which are found in natural scenes, as shown in Fig.~\ref{f2}A-E. To span these ranges, contrast values $C$ had to be chosen in the low range, due to the hardware limitations of the stimulus display. 

To quantify how retinal neurons change their code when  contrast, skewness, or kurtosis of the stimulus change, we constructed accurate encoding models for the recorded neurons, and compared their properties under the different stimuli. We thus followed \citet{fairhall06}, who have shown that for spatially uniform Gaussian stimuli in the salamander retina, linear-nonlinear (LN) models with one or two linear filters often suffice to describe the cells' encoding scheme with high accuracy. Moreover, \citet{baa1} and \citet{rdr} also provided an interpretation of the filtering operations as dimensionality reduction on the stimulus space, the success of which can be quantified with information theory. Here we extended their framework to non-Gaussian stimuli and analyzed how information is encoded beyond linear filtering stage, in the nonlinear response, and finally in the spiking rate. We could then characterize adaptation and invariance quantitatively, and  compare the behavior of real neurons with computational models that either have or lack adaptation.

\subsubsection{Linear filters of retinal ganglion cells do not adapt to changes in higher-order stimulus statistics }

\begin{figure}[tb]
\includegraphics[width=3.5in]{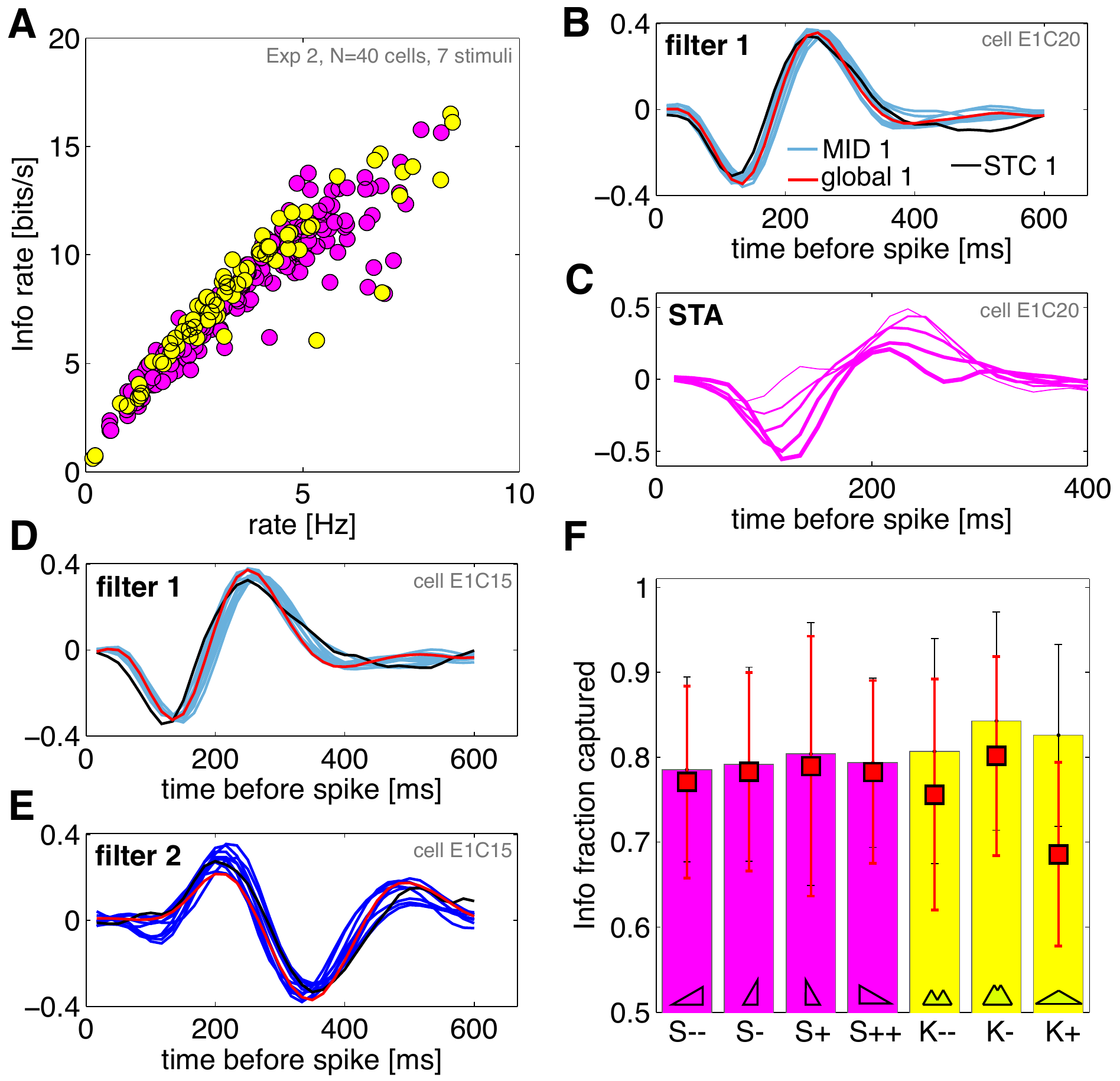}
\caption{{\bf Linear filters and higher-order statistics stimuli in retinal ganglion cells.} {\bf A)} Estimated information rate (see Methods) as a function of the mean firing rate. Each dot represents one of the 40 cells in experiment 2 exposed to one of the 7 HOS conditions (skewed stimuli in magenta, kurtotic in yellow). The growth in information is slightly sublinear, with no obvious systematic dependence on the stimulus type. {\bf B)} A cell whose behavior is captured well by a single linear filter. Shown in light blue are the filters for all 9 (2 Gaussian, 7 HOS) stimuli reconstructed using maximally informative dimensions; in black the spike-triggered average computed on the Gaussian stimulus C++; in red, a single \emph{global} filter inferred using MID across all stimulus conditions simultaneously. {\bf C)} Biased STA filter estimates for 5 skewed stimuli (thicker lines mean increasing skewness) for the same cell as in B (note the difference in the time axis). {\bf D,E)} A cell whose behavior is described well by two linear filters (light blue = the most informative dimension; dark blue = the second most-informative dimension). Other symbols the same as in B). {\bf F)} Information captured by two filters (across stimuli, horizontal axis), as a fraction of the total information per spike; mean and std across 40 cells in experiment 2. The average performance of global models (the same pair of filters across all stimuli for each cell) is plotted as red squares.   }
\label{f5}
\end{figure}

We recorded from 23 retinal ganglion cells that were presented with 9 types of non-repeated stimuli (2 Gaussian + 7 higher-order statistics) in experiment 1, and from 40 cells presented with non-repeated and repeated  stimuli of 7 types with higher-order statistics  (see Materials and Methods) in experiment 2.  Figure~\ref{f5}A shows the estimated information rate of the neurons as a function of their firing rate. Consistently with previous reports \citep{vijay09}, the information rate $I$ scaled weakly sub-linearly with the mean firing rate $\bar{r}$ ($I\propto \bar{r}^{0.76}$). There were no other large systematic dependencies in transmitted information across  cells and stimulus classes.

Next, we inferred the best linear filters for each cell, and each of the stimulus condition $\mathcal{S}$, separately. We used maximally informative dimensions (MID) for learning the filters for all stimuli, and for the Gaussian stimuli we additionally used spike-triggered average and spike-triggered covariance. We also inferred a \emph{global} model for each cell, where a single filter (or a single pair of filters) was fit across all stimulus conditions (see Materials and Methods). In cross-validation on test data, the prediction performance of the models of essentially all cells ($97\%$ of cell/stimulus combinations) increased when using two filters (2D LN models), compared to one-dimensional LN models, but in some cases the contribution of the second filter was very small. The linear filters inferred using MID for one of these cells are shown in Fig.~\ref{f5}B for 9 stimulus conditions; overlaid is the leading eigenvector of the spike-triggered covariance matrix computed for the C++ stimulus, and the best global filter learned by MID (all filters are scaled to unit norm). The filters  show very strong overlap, indicating that their shape does not adapt to the stimulus distribution. We emphasize that computing the naive STA estimates  gives a systematic change in filter shape with the stimulus skew, as shown in Fig.~\ref{f5}C, but this is simply an artifact of the STA estimation on non-spherically-symmetric stimuli, and is not indicative of any adaptation process.

Figures~\ref{f5}D-E show a typical cell for which a model with two linear filters is needed. We again observe a high overlap between the filters inferred using MID in 9 stimulus conditions, the filter pair computed using STC in the Gaussian condition, and the single global best pair of filters inferred across all conditions using MID. 15 of the 23 cells in experiment 1 have an average firing rate above $1.5\e{Hz}$ for every stimulus, permitting reliable filter estimation. Out of those, a single-filter model in the Gaussian condition suffices for 8 cells (i.e. the single-filter model accounts for more than 90\% of the information per spike of the two-filter model), and for 7 cells two filters are needed. To measure the agreement between inferred filters across conditions for each cell, we compute the Pearson cross-correlation between the STC derived filter(s) in the Gaussian condition, and the filters derived using MID for each stimulus condition $\mathrm{S}$. The average correlation across the group of cells with a single linear filter is $97\%\pm 2\%$, while the average correlation across the group of cells with 2 filters is $86\%\pm 5\%$ (error bar = std across cells); the decrease in the later case is mainly attributable to the difficulty of inferring jointly 2 filters using MID with a limited number of spikes.

\begin{figure}[bt]
\includegraphics[width=3.5in]{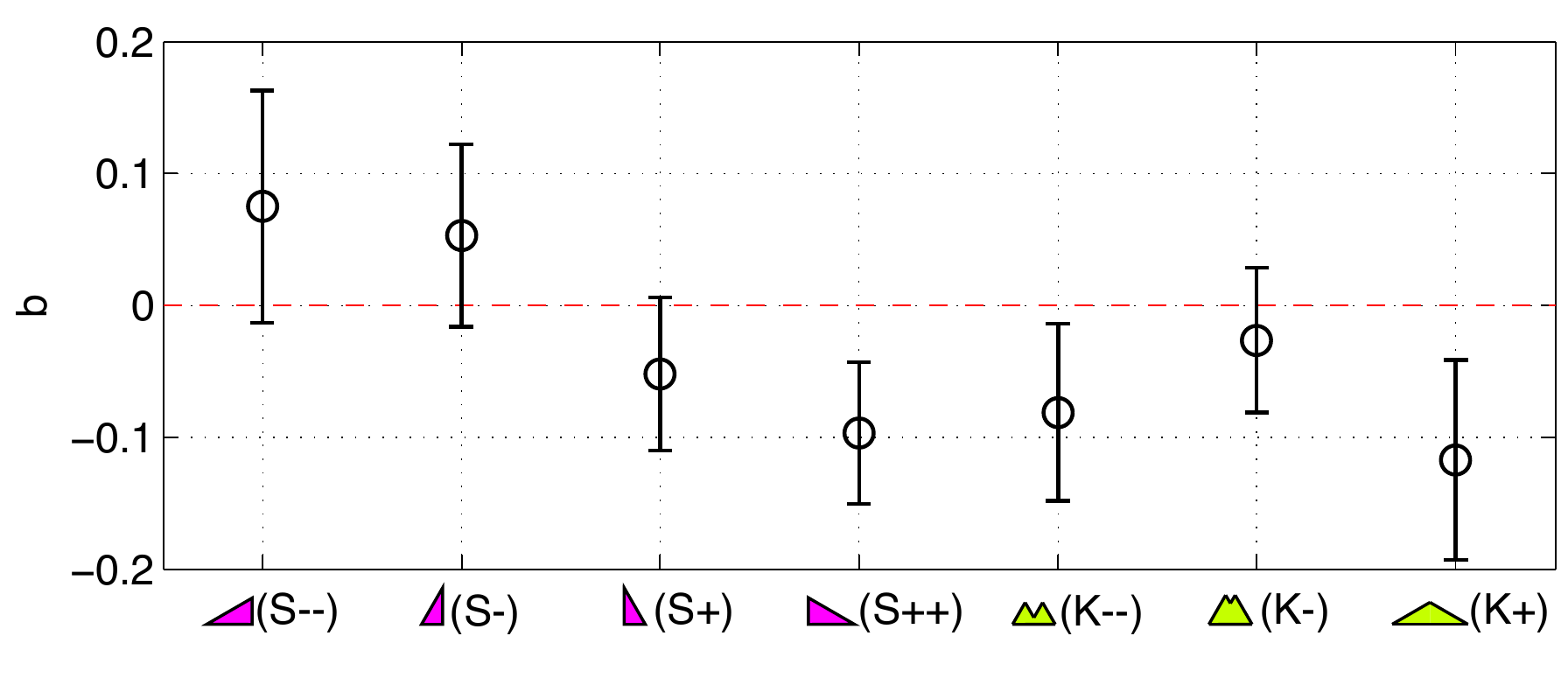}
\caption{{\bf The dependence of the  balance index on the stimulus type}. The balance index $b$ is a ratio between the total (signed) area under the filter, normalized by the absolute area; balanced filters have $b=0$, fully unbalanced $b=\pm 1$. Shown are averages ($\pm 1$ std error bars) for 40 cells in experiment 2, computed across both linear filters for each cell.}
\label{f6}
\end{figure}

We quantified the effects of changing the higher-order statistics on the shape of the linear filters by computing the balance index $b$, defined as the ratio between the total (signed) area under the filter and the absolute area: $b=\sum_{\mu} k_{\mu} / \sum_{\mu} | k_{\mu} |$, where $\mu$ indexes the temporal components of the filter. 
The balance index across the recorded population was $\bar{b}=-0.0350\pm 0.13$ (n=40 cells from experiment 2), where the mean and error bar (std) are taken across all cells, conditions and both filters. Broken down across conditions, there is a small systematic modulation of $b$ with the stimulus, which is smaller than 10\% (Fig~\ref{f6}); additionally, there is  substantial scatter across the recorded population. 

To ask whether these slight variations in filter shape across stimuli matter for encoding, we compared the performance of global filters (constant for each cell across all the stimuli) with filters inferred separately for each stimulus. We estimated the information captured by the single-filter model, by a two-filter model, and by a two-filter global model (where a single pair of two filters is inferred for all stimuli for each cell). 
Single-filter models (fit to each stimulus separately) captured $69\% \pm 12 \%$ of the information  per spike (averaged across cells and stimuli). Two-filter models (fit to each stimulus separately) captured $81\% \pm 12\%$ of information per spike, as shown in Fig.~\ref{f5}F; for some cells, two filters capture essentially all of the information. Our observations were quantitatively consistent with the results reported by \citet{fairhall06}. Most importantly, the global models where filters did not change with the stimulus capture $77\% \pm 13\%$ of the information per spike, confirming that the minor variations between filters inferred in different stimulus conditions are insignificant from the encoding perspective. 

Taken together, our results show that the shape of linear filters in 2D LN models for salamander retinal ganglion cells does not change noticeably when the skewness and kurtosis vary across the range observed in natural scenes. 

\subsubsection{Nonlinearities of retinal ganglion cells responding to higher-order statistics stimuli}

\begin{figure*}[bt]
\includegraphics[width=5in]{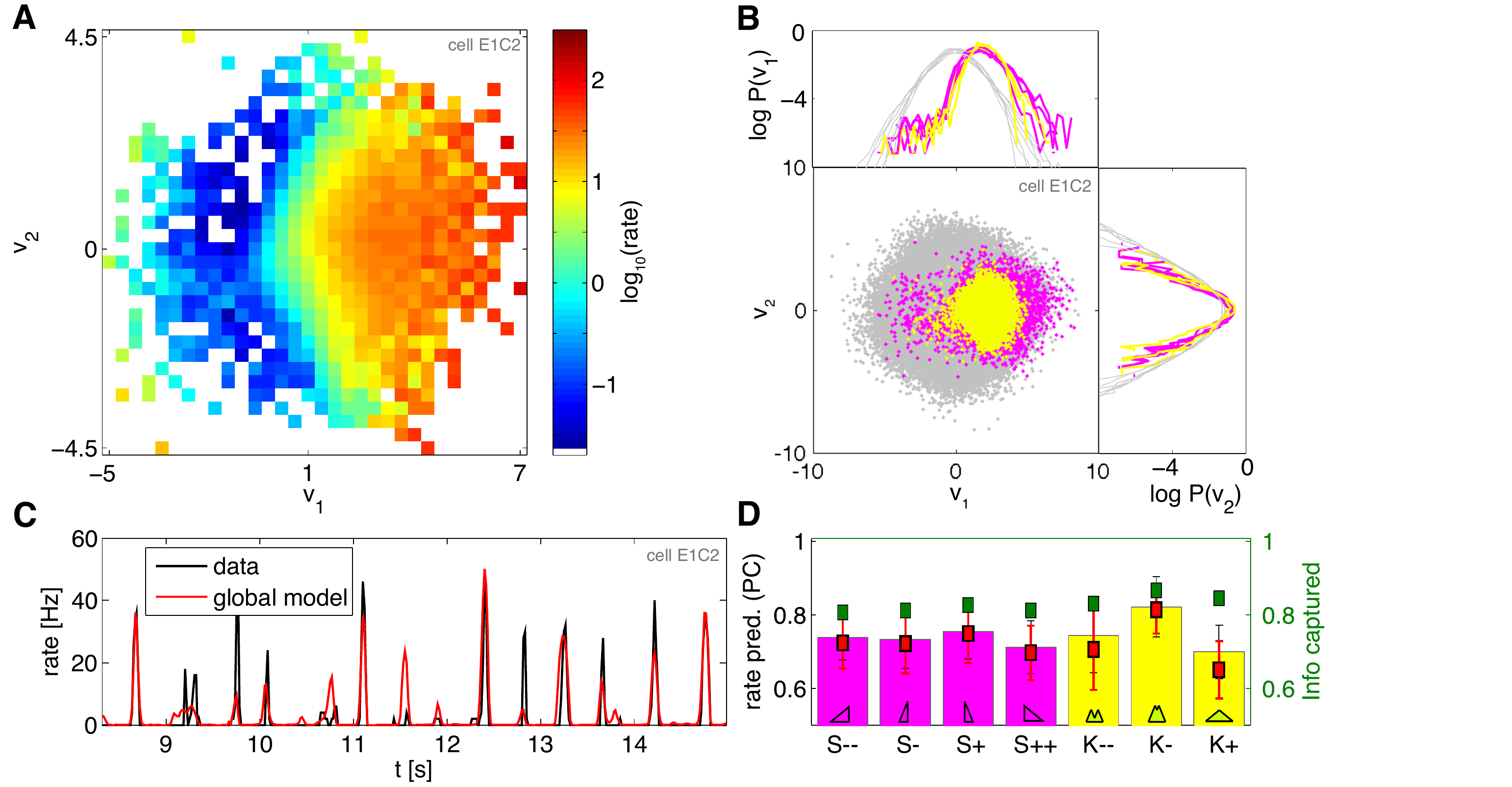}
\caption{{\bf Nonlinearities and rate prediction with higher-order statistics stimuli.} {\bf A)} A 2D nonlinearity globally fit across all HOS stimuli for  neuron E1C2; projection of the first (second) filter shown on the horizontal (vertical) axis. Hotter colors indicate increased firing rates (see colorbar, rate in Hz; white = regions of $\{v_1,v_2\}$ space where no spike or prior samples have been observed). {\bf B)} For the same cell, the depiction of prior ensemble (gray dots, all 7 higher-order statistics stimuli overlaid) and the spike-triggered ensembles (magenta = skewed stimuli, yellow = kurtotic stimuli); shown are also projections of the data, i.e. the marginal distributions  $P(v_1)$ and $P(v_2)$, on the logarithmic scale, for all 7 stimuli separately. {\bf C)} The segment of predicted and true firing rate in responses to repeated K- stimulus presentations (red = 2D LN global model fit to all stimuli for this neuron; black = true rate). {\bf D)} Model performance, measured as the Pearson correlation (PC) between the true and predicted PSTH, across different stimuli (horizontal axis; average and error bars = mean and std across 40 neurons in experiment 2). The performance of 2D LN models fit separately for each stimulus is shown by magenta (skewed stimuli) and yellow (kurtotic stimuli) bars. Global model performance (red squares) matches the performance of models fit separately. Right axis, in green: information fraction captured by the nonlinear combination of the 2 linear projections, $\mathcal{N}(v_1,v_2)$, shows no drop compared to the information captured by the linear features themselves (c.f. bars in Fig.~\ref{f5}F), and is between $80-85\%$ across all stimuli (error bars omitted for clarity, comparable to error bars in information fraction captured by the 2 linear features).}
\label{f7}
\end{figure*}

Completing the LN description of the ganglion cells is the mapping from the linear projection(s) of the stimulus into the cell's firing rate. We estimated these 2D nonlinear functions from the data by binning $P(v_1,v_2|{\rm spike})$, where $v_i=\mathbf{k}_i\cdot \mathbf{s}$ are the projections of the stimulus onto the two filters, $\mathbf{k}_1,\mathbf{k}_2$, as explained in Materials and Methods. This was done for each neuron and each condition separately, or for all conditions jointly using the \emph{global} pair of filters, to yield a single 2D global LN model for every cell. Figure~\ref{f7}A shows a global nonlinear function for an example neuron. In Fig.~\ref{f7}B we explicitly show, for that same neuron, the prior ensembles for all 7 higher-order statistics stimuli (gray), with overlaid spike-triggered ensembles for skewed (magenta) and kurtotic (yellow) stimuli, along with the marginal projections of these distributions. The number of spikes is in general insufficient to reliably sample and then systematically compare the 2D nonlinear functions across cells and stimuli. We therefore decided to base our comparisons directly on the firing rate prediction performance. The 2D models for cells in experiment 2 are fit on the non-repeated segments, and are subsequently tested by predicting PSTH in response to repeated stimulus presentations for which we could measure the true PSTH; here, too, the prediction was done either with models fit separately at each condition, or with the global model, where 2 linear filters and the nonlinearity were fit simultaneously across all stimuli, as shown in Fig.~\ref{f7}C. 

Does the nonlinearity change with the stimulus condition? We first estimated how much information is lost in compressing the 2D projections $\{v_1,v_2\}$ into the nonlinear combination, $\mathcal{N}(v_1,v_2)$, on average. As shown in Fig.~\ref{f7}D, the nonlinearity captured 80-85\% of the information per spike (fit for each condition separately), essentially the same amount as the two linear filters (c.f. Fig.~\ref{f5}F). This finding establishes that the nonlinear mapping itself does not discard the information per spike, and that analyzing the changes in point-wise nonlinearities is warranted. In terms of PSTH prediction, the prediction of the 1D LN models, fitted to each conditions separately, had $64\%\pm 8\%$ correlation with the real PSTH (error bar = std across 40 cells and 7 stimuli). 2D LN models were better with $74\%\pm 9\%$, as shown in Fig~\ref{f7}D. The global models that had constant filters and nonlinearity across 7 higher-order statistics conditions, performed negligibly lower, with $72\%\pm 9\%$ correlation, mostly due to the sparse K+ stimulus condition (where the spike rate is the lowest); there was essentially no drop for other stimulus conditions in performance of global vs. separate models. Our simulations show that the limitations to the performance of LN-based models in predicting the firing rate are due to the inability of these models to capture the refractory and spike-feedback effects (which would be possible with e.g. generalized linear models \citep{pillow08} or Keat-type models \citep{keat01}). We reiterate, however, that the prediction performance  was used here solely as a differential measure, to evaluate the significance of the changes in linear filters and nonlinearities with the stimulus condition.
\begin{figure}[bt]
\includegraphics[width=3.3in]{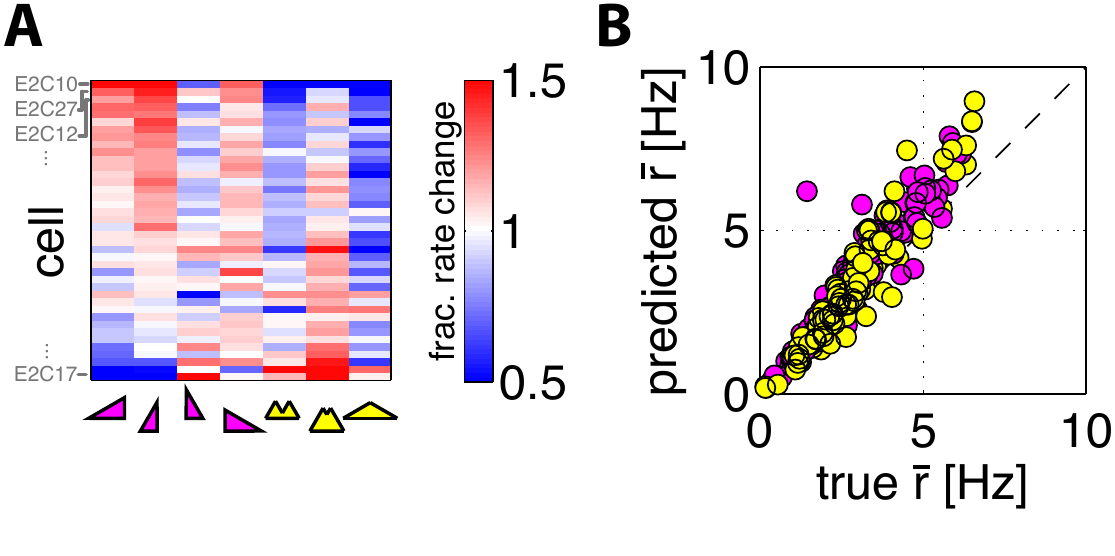}
\caption{{\bf Measurement and prediction of  changes in the mean firing rate with stimulus condition.}
{\bf A)} The relative change (color scale, 1 = the firing rate of the cell is equal to the mean rate for that cell across all stimuli) in the mean firing rate for 40 cells of experiment 2, as a function of the stimulus condition (magenta = 4 skewed, yellow = 3 kurtotic stimuli). Neurons (rows) were sorted by the projection on the first principal component explaining most of the change across the recorded population; cells close to the top increase the firing rate in response to left-skewed stimuli, while cells at the bottom increase the rate in response to right-skewed and negative kurtosis stimuli. {\bf B)} Global 2D models for each cell predict the average rate well (each dot is one cell in one of the 7 HOS stimulus conditions).}
\label{f7bis}
\end{figure}

Finally, we can ask how successfully the global models recapitulate the overall firing rate changes with the stimulus statistics. Figure~\ref{f7bis}A shows the relative change in firing rates for cells from experiment 2 for 7 HOS stimuli; the cells have been sorted to reveal the dominant pattern, where cells that prefer left-skewed stimuli respond less strongly to the other stimuli, while cells that respond strongly to right-skewed stimuli also respond to negative kurtosis. We can use a global 2D LN model fit for every cell to predict the mean firing rate in response to all of the 7 stimuli. These models (without any adaptation) reproduced very well the mean firing rate for each stimulus and cell, as depicted in Fig.~\ref{f7bis}B, and therefore also the pattern of changes in the firing rate (not shown).
\subsubsection{Adaptation to changes in contrast and invariance to changes in higher-order statistics}
We observe that the encoding properties of salamander ganglion cells do not depend on the higher-order statistics of the luminance levels. To establish this, we compared the shape of linear filters across the stimulus conditions directly (by measuring their overlap), by information-theoretic measures (information per spike captured), and through the impact on the prediction performance; similarly, we assessed the changes in the nonlinearity by measuring their impact on the prediction performance. In all cases---with the possible exception of highly kurtotic (K+) stimulus---we found that global models, i.e. models with invariant pair of filters and invariant nonlinearity, account for the neural behavior equally well as the models fit to different stimuli separately. Since neurons are nonlinear dynamical systems coupled to the stimulus, we find this extent of invariance surprising -- we expected that higher-order statistics in the stimuli would ``interact'' with the photoreceptor-bipolar-RGC cascade to yield a notable change in the effective encoding properties, especially for heavily skewed or bimodal stimuli. 

\begin{figure*}[!t]
\includegraphics[width=6in]{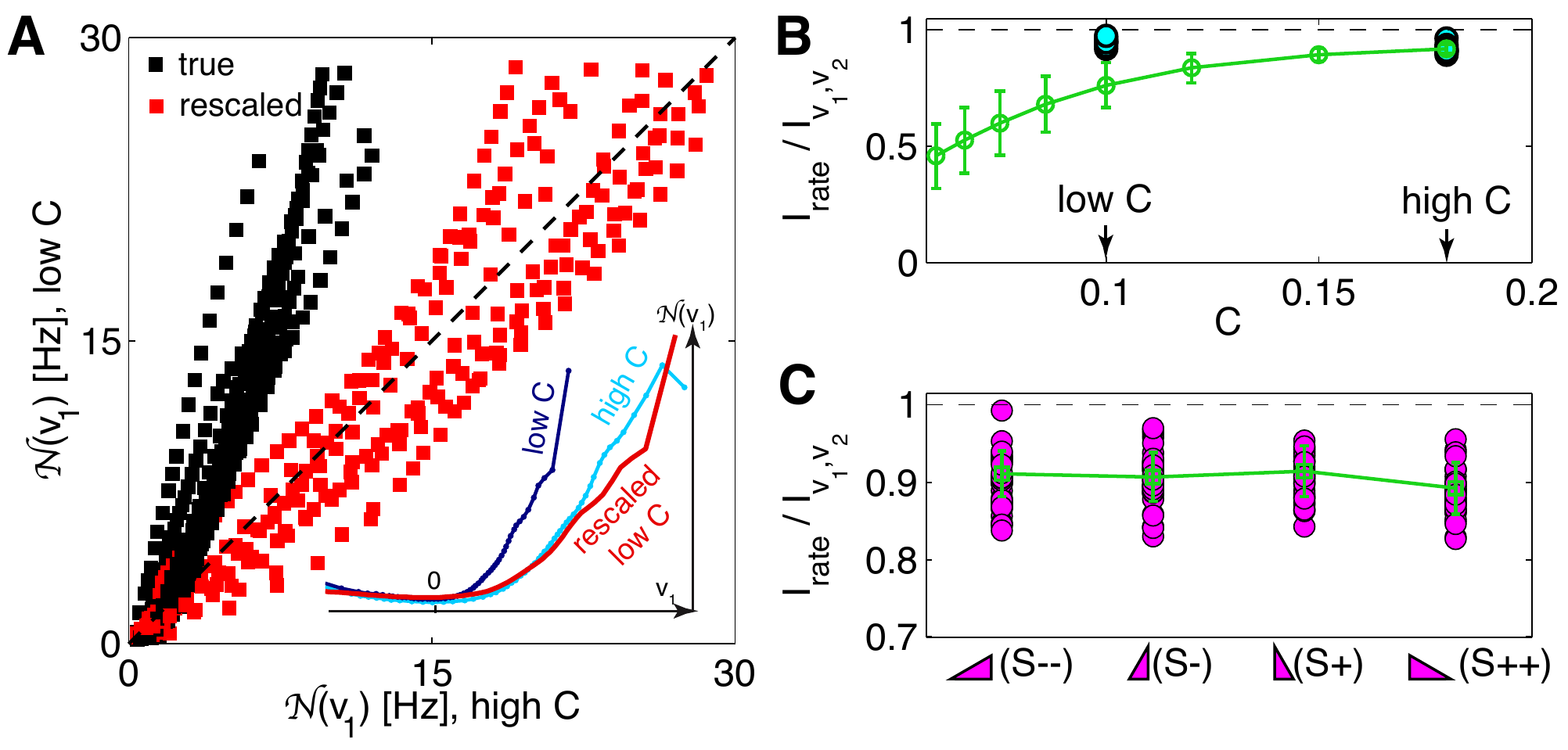}
\caption{{\bf The benefits of contrast and higher-order statistics adaptation.} {\bf A)} \emph{Inset.} The 1D nonlinearity, $\mathcal{N}(v_1)$, for an example neuron (E0C4) inferred at high contrast (C++, light blue), and at low contrast (C+, dark blue). The low contrast nonlinearity can be aligned to the high contrast one by (i) rescaling the stimulus (horizontal) axis by the ratio of the two contrasts, and (ii) rescaling the firing rate (vertical) axis by the ratio of the two average firing rates, yielding the red line. \emph{Main panel.} Scatter plot of the nonlinearity at high vs nonlinearity at low contrast (black, 19 neurons from experiment 1; the coordinates of each point are the high / low C  nonlinearity values at the same value of the projection $v_1$ for a particular neuron). After rescaling, the nonlinearities align (red). The scaling breaks down for rates above $30\e{Hz}$ (rarely observed at low contrast). {\bf B}) The information in the spiking pattern of a LN model neuron, normalized by the information captured by the two linear projections of the corresponding stimulus. Cyan circles = inferred models for 19 neurons for high and low contrast (C++, C+) stimulus. Green line = computational prediction obtained by taking 19 high contrast models and dialing down the stimulus contrast without any adaptation in the model (error bars = std across the neurons). {\bf C)} Analogous analysis for changes  in skewness (note the difference in scale); magenta =  models inferred separately for each skewed stimulus; green = invariant (and therefore non-adapting) global models for every cell. }
\label{f8}
\end{figure*}

We next analyzed the recordings from experiment 1 where neurons were exposed to high (C++) and low (C+) contrast stimuli. Since our analysis kept the filters normalized to unit length, contrast adaptation would be reflected in the change of the shape of the nonlinearity. Indeed, this can be seen in Fig.~\ref{f8}A (inset), which shows the (marginal) nonlinearity, $\mathcal{N}(v_1)$, of a typical neuron for the high- and low-contrast experiments. We then took the nonlinearity from the low-contrast experiment and rescaled it as follows. First, we rescaled the range of inputs to the nonlinearity, $v_1=\mathbf{k}_1\cdot\mathbf{s}$, by the ratio of high to low contrast, C++/C+$\, = 1.82$. Second, we also rescaled the output firing rate by the ratios of the steady-state firing rates in both C+ and C++ conditions (the rates are not equal because the neurons do not adapt perfectly). After these two rescaling operations, the \emph{measured} nonlinearity for C++ (high contrast) stimulus lined up well with the \emph{rescaled} nonlinearity from the C+ (low contrast) stimulus, indicating the ability of the neuron to adapt to contrast. This observation was generally true for most (19 out of 23)  neurons in our dataset, as shown in Fig.~\ref{f8}A. The rescaling fails at very high firing rates, because they are not accessed in the low contrast condition, and (potentially) because we were only looking at the marginal 1D (and not the full 2D) nonlinearity.

When the stimulus contrast changes, retinal ganglion cells adjust their gain, matching the variation of the signal about the mean to the dynamic range of the firing rate at the output, thereby keeping the information rate high. Without adaptation, the information rate would drop because the neurons have a limited output range and they are noisy. ``Noise'' in the context of LN encoding models is the stochasticity related to the spike generation: from the stimulus $\mathbf{s}$ to spike, $\mathbf{s}\rightarrow\{v_1,v_2\}\rightarrow\mathcal{N}(v_1,v_2)\rightarrow \mathrm{spike}$, it arises when the nonlinear function $\mathcal{N}$ is interpreted as the mean firing rate of  a  Poisson point process. To  explore the effects of the presence or absence of adaptation, we  generated spikes according to this LN prescription in response to various stimuli, and measured the information in such synthetic spike trains using Eq.~(\ref{brenner}). By using the true inferred (adapting) models for low and high contrast, we found that in both conditions the real spiking neuron can retain $\sim 90\%$ of the information extracted from the stimuli by the linear filters. On the other hand, when using the model inferred at high contrast, holding it fixed (no adaptation), and probing it with stimuli of progressively lower contrast, the information rate dropped significantly, as shown in Fig.~\ref{f8}B. The situation is very different for skewness (or kurtosis):  Fig.~\ref{f8}C shows that no such drop in information is observed when the global model is used to generate spikes in case of skewed stimuli, making adaptation unnecessary and invariant encoding possible.

\begin{figure*}[!t]
\includegraphics[width=6in]{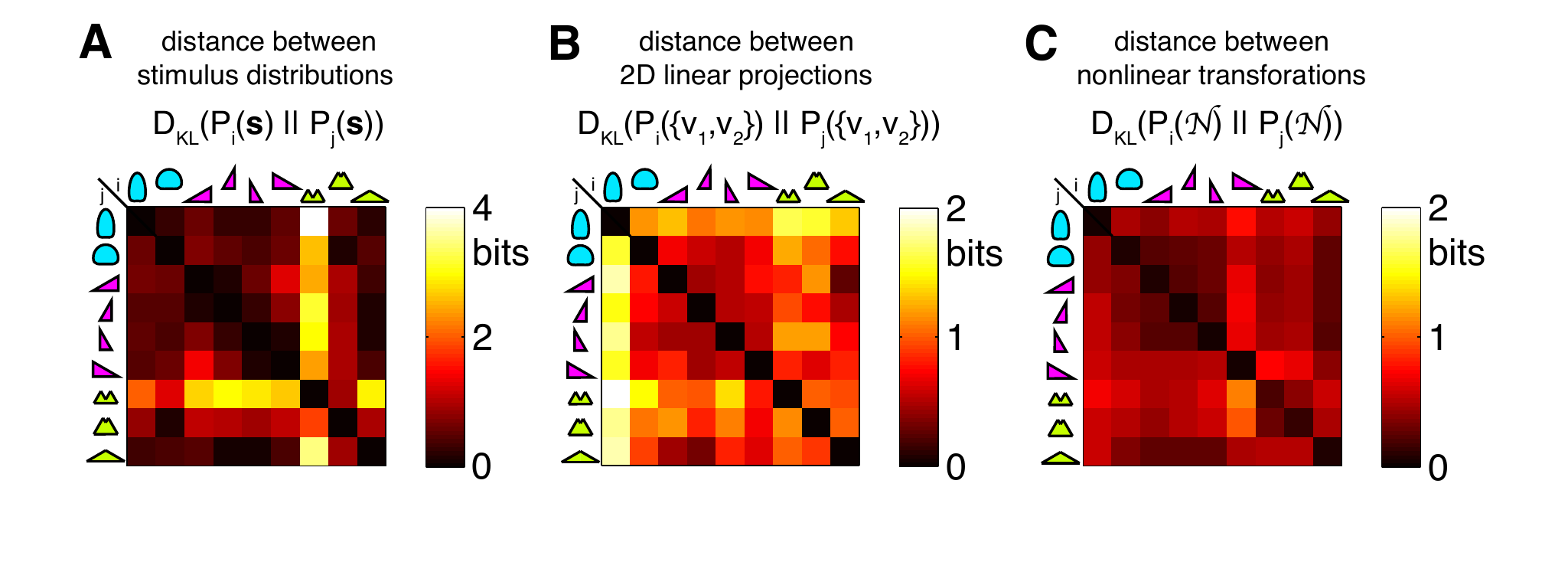}
\caption{{\bf Neurons with contrast adaptation yield similar distributions of firing rates in response to very different distributions of inputs.} {\bf A)} Kullback-Leibler distance matrix, $D_{KL}$ (bits), between all 9 pairs of stimulus distributions (cyan = 2 Gaussian, magenta = 4 skewed, yellow = 3 kurtotic distributions). Bimodal K{-}{-} distribution is most different from the others. {\bf B)} The difference between the respective 2D linear projections of the 9 stimulus distributions (shown are the averages over $D_{KL}$ matrices for 19 neurons in experiment 1). Linear filtering of IID stimuli washes out most of the higher-order structure (but not the second order), and the most distinct stimulus type at this stage is C+, since its variance is different from the other distributions of projections. {\bf C)} $D_{KL}$ (average over 19 neurons) between the nonlinear transformations of the respective linear projections. Since the nonlinearity adapts to  contrast, this step equalizes the low contrast (C+)  with the other stimuli.}
\label{f9}
\end{figure*}

This effect is easy to understand if we compare the extent to which the 9 stimulus distributions differ a priori, after filtering by the neuron's linear filters, and after passing through the nonlinear function. Figure~\ref{f9}A  shows a $9\times 9$  matrix of the Kullback-Leibler distances $D_{KL}(P_i(s) || P_j(s))$ between all pairs of stimuli $i,j=\{$C+,C++,S{-}{-},S-,S+,S++,K{-}{-},K-,K+$\}$ (2 Gaussian, 7 higher-order statistics). The bimodal stimulus K{-}{-} is clearly distinct from the others. After linear filtering (Fig.~\ref{f9}B), however, the low contrast stimulus C+ differs the most from the others, which are all matched in contrast. Because  linear filters, as we have shown, do not change in shape, this is simply a consequence of the central limit theorem: the filters sum up (with weights) samples drawn independently from the stimulus distributions $P_{\mathcal{S}}$, so the filter outputs must converge to Gaussian distributions with variances that are related to the variance (or contrast) of the input. In other words, the invariant linear filters remove the signatures of higher order statistics and ``equalize'' different stimuli with the exception of their contrast. In the last, nonlinear stage (Fig.~\ref{f9}C), the nonlinearity adapts to contrast as well, ultimately yielding LN model outputs whose distributions are very similar across the range of stimuli  differing in contrast, skewness and kurtosis.
\subsubsection{Should the linear filters adapt to changes in higher-order statistics?}
We previously established that the retinal ganglion cells do not adapt their linear filtering properties to changes in stimulus skewness and kurtosis. Taking this invariance of linear filters as given, we next showed that linear filtering and contrast adaptation together can successfully remove the signatures of higher-order statistics from the distribution of neural firing rates. Here we return  to the observation of invariant linear filters to ask whether such invariance should be expected on theoretical grounds.

In a one-dimensional LN model neuron, the probability of spiking  was assumed to be a saturating nonlinear function of the filtered stimulus:
\begin{equation}
P(\mathrm{spike} | \mathbf{s}) = \frac{1}{2}\tanh(\mathbf{k}(\mathcal{S})\cdot \mathbf{s} + \theta(\mathcal{S})) + \frac{1}{2}, \label{model}
\end{equation}
where the linear filter $\mathbf{k}$ and the spiking threshold $\theta$ may depend on the stimulus type $\mathcal{S}$. We  simulated spike sequences $\sigma$ of this model neuron in response to repeated presentations of the stimulus, whose value in each time bin was drawn independently from from $P_{\mathcal{S}}(L)$: $\sigma(t,r)=\{0,1\}$ was `0' when the neuron was silent in time bin $t$ and during repeated presentation $r$ of the stimulus, and `1' if it spiked. To quantify how well such a neuron encodes information about the stimulus into  spike trains, we estimated the information rate $I$ (in bits per second) between the stimuli and the response, using the direct method \citep{strong98}.

\begin{figure*}[bt]
\includegraphics[width=7in]{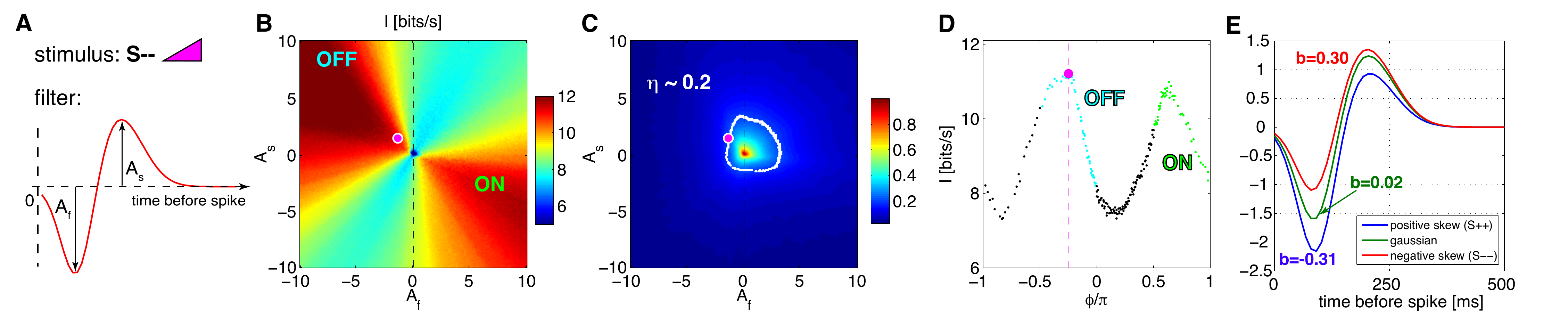}
\caption{{\bf Finding optimal filters for an LN model neuron for stimuli with negative skew (S{-}{-})}. {\bf A)} The biphasic filter with two parameters determining the amplitudes of the fast and slow lobes, $\{A_f,A_s\}$. Each of the amplitudes can be positive or negative. When $A_s$ is positive and $A_f$ is negative, the cell is an OFF cell; when $A_s$ is negative and $A_f$ is positive, the cell is an ON cell. {\bf B)} Information about the stimulus encoded in the spike train (bits per second, color scale), as a function of the fast lobe amplitude $A_f$ and the slow lobe amplitude $A_s$. The ON and OFF types have been denoted in the 2 corresponding quadrants of the plot. {\bf C)} Fraction of entropy lost to noise, $\eta$, as a function of $\{A_f, A_s\}$. Points in the plane that have $\eta\sim 0.2$ are shown in white, and lie on a circular locus of points; we are only interested in the models with fixed value of $\eta$. We  parametrize such points by their angle, $\phi$, going counterclockwise from the vertical ($A_f=0$). {\bf D)} Information on the locus of points $\eta\sim 0.2$ as a function of $\phi$; these values are extracted from B) along the $\eta=0.2$ contour. Green points correspond to ON cells, cyan points to OFF cells. There are two peaks in information, one (slightly higher) peak for the OFF type cell and one for the ON type cell. In all plots, the optimal OFF cell is denoted by a magenta dot. {\bf E)} Theoretical prediction for the shape of the optimal biphasic OFF filters for stimuli with different skewness values. As  skewness increases from negative (S{-}{-}, red) to positive (S++, blue),  the negative lobe becomes more prominent and the positive lobe becomes less prominent. For the symmetric Gaussian stimulus (C++, green) the optimal filter is balanced. These changes are quantified by the balance index $b$ (see text), which measures the difference in area between the lobes, normalized to the total absolute area under the filter. For the simulations in this figure, the stimulus refresh time is $10\e{ms}$ and mean firing rate is held fixed at $5\e{Hz}$ (results are qualitatively unchanged for rates up to fourfold higher). }
\label{f3}
\end{figure*}

For stimulus types $\mathcal{S}$ of varying skewness, we found the optimal filter $\mathbf{k}$ and the threshold $\theta$  that would maximize the information rate $I$ that the neuron would convey. If real neurons were adapting in such a way, this procedure would then predict how their filters would change with the stimulus distribution. For computational reasons, our model was simplified: {\bf (i)} the linear filter was biphasic, with a ``fast'' lobe of amplitude $A_f$, and a ``slow'' lobe of amplitude $A_s$, whose widths and the positions were fixed, so the filter was fully specified by the two amplitude parameters $\{A_s, A_f\}$ as schematized in Fig.~\ref{f3}A;  {\bf (ii)} we maximized  the information rate $I$ for a fixed average firing rate $\bar{r}$;  {\bf (iii)} the effective noise of the neuron, or  fraction of output entropy that is lost  to noise, $\eta=S_{\rm noise}/S_{\rm total}$, was fixed. 

Figure~\ref{f3} shows the dependence of the information rate on the shape of the stimulus filter: left-skew distributions (S{-}{-}) slightly favor OFF cells (negative fast lobe, positive slow lobe, Fig.~\ref{f3}B), while right-skew distributions (S++) slightly favor ON cells (not shown). This  conclusion was robust to noise in the neuron $\eta$ ranging from $\sim 0$ to  0.5. For a specific choice of $\eta = 0.2$ in Fig.~\ref{f3}C, Fig.~\ref{f3}D shows the  dependence of information rate on the ratio of the fast to slow lobe amplitude, $\tan\phi = A_f/A_s$. For each stimulus (S++ and S{-}{-}), there are two local maxima, one for an ON-like and one for an OFF-like cell, which differ in the transmitted information by less than 10\%. Importantly, however, the maxima are not achieved at the same value of $\phi$ in both stimulus conditions -- this means that while an adapting ON or OFF cell can maintain the same rate of information transmission when the stimulus changes, it will need to modify the filter shape by adjusting the ratio $A_f/A_s$.

Focusing on the case of an OFF cell, we found that an optimally adapting cell would increase the area under the fast (negative) lobe and would decrease the area under the slow (positive) lobe with increasing skewness.  We used the  balance ratio $b$, introduced previously, to quantify this change.
Figure~\ref{f3}E shows significant changes in the filter shape with skew that range from $b=0.3$ for S{-}{-} to $b=-0.3$ for S++ (as expected, the optimal filter for the Gaussian stimulus is balanced ($b=0$), with equal and opposite areas under the two lobes). However, these substantial changes in the filter shape only lead to moderate changes in the amount of encoded information: in the case shown in Fig.~\ref{f3}, the information gain of the adaptive neuron relative to the case of no adaptation is always less than 10\%. While the exact number varies with the chosen constraints ($\bar{r}$, $\eta$, the locations and widths of the filter lobes), two qualitative observations remain true: {\bf (i)}  the neurons should adjust the biphasic linear filter away from the balanced configuration (optimal for Gaussian stimuli) by systematically adjusting the weight under the fast and slow lobes so that, e.g. in case of the OFF cell, negative skew favors larger weight in the slow lobe; but also that {\bf (ii)} these adaptive changes would only lead to  small relative information gains.

\section{Discussion}

We explored the limits of retinal adaptation, by analyzing the responses of salamander retinal ganglion cells to temporally uncorrelated and spatially uniform stimuli with Gaussian, skewed and kurtotic luminance distributions. While the retina is highly adaptive to changes in first and second order statistics, we found invariance of neural encoding to changes in stimulus skewness and kurtosis in both the linear filtering stage and the nonlinear stage. \citet{bonin08} reported a similar result in the cat LGN for spatially structured stimuli, using 1D LN models inferred using reverse correlation; we see an analogous invariance to skew and kurtosis for spatially uniform stimuli in the retina, for 2D models inferred using a theoretically unbiased method, while exploring a substantially wider range of skewness and kurtosis values. Taken together, these results suggest that adaptation in the early visual system is not sensitive to changes in commonly measured higher-order statistics.

From an information optimization point of view it is not immediately clear why the retina does not adapt to such higher-order changes. Moreover, since neurons are nonlinear dynamical systems coupled to their inputs, a large change in the statistical structure of the input would, on general grounds,  be expected to influence the neuron's effective encoding properties, even if it did not lead to an increase in information transmission. Thus, invariant encoding in the face of substantial changes in the stimulus statistics is not an obvious a priori expectation.

Adaptive code is necessarily ambiguous. The same response from an adapting neuron can, for example, signal two different light intensities, depending on the stimulus history. Downstream neurons must therefore rely either on keeping track of the adaptive state of the encoding neuron, or on using diversity in the neural population in a proper way to estimate the stimulus. Moreover, nontrivial processing is required also on the encoding side: the adaptive neuron needs to infer from the stimulus itself whether some underlying property of the stimulus (such as the mean luminance or contrast) has changed and thus an adaptive response needs to be triggered (but c.f. also \citet{borst05}). In short, adaptive codes incur computational costs that invariant codes do not.

These considerations suggest that there exist changes in stimulus statistics for which the benefits of adaptation outweigh the costs. It also implies the existence of changes where the encoding properties \emph{should} remain invariant, even if other equally good encoding strategies were available to the cell, thereby making the code easy and unambiguous to decode.  Such changes would be those where, even without adaptation, the information rate would not drop precipitously when the stimulus distribution changes. 

Our results lead us to believe that an encoding model with non-adapting linear filter(s) and a nonlinearity, which adapts to contrast but not to higher-order statistics, is an efficient  strategy that reduces the inherent ambiguity in the code. Contrast adaptation is necessary to prevent a severe drop in the information rate after a contrast change. If in addition the linear filters are invariant, the filtering, together with the central limit theorem, can remove higher-order statistics. This would  leave the luminance and the contrast as the only target statistics for adaptation, while simultaneously maintaining a high information rate and minimizing the code ambiguity. 

We did not observe any adaptive changes in the filter shape either for skewed and kurtotic stimuli, or  after the change in contrast. Small changes in filter shape have been previously reported by \citet{smirnakis97} in response to large changes in contrast, beyond the values we used here (``faster'' filters at high contrast values). It is not clear how one could assess the  cost of adding  an adaptive mechanism that modifies the retinal ganglion cell filters for higher-order statistics; we  can ask, however,  how much  information would be gained by it. Our toy model simulations show that while it would be theoretically possible to increase the transmitted information by adjusting the shape of the linear filter to the stimulus, the gains in information would be very modest (around 10\% for a robust range of realistic parameter choices, much less than the gain due to contrast adaptation), despite significant changes in the shape of the optimal linear filter.

One explanation for the lack of dynamic adaptation to higher-order statistics is therefore that it does not yield much gain in information, while potentially increasing the coding cost and complexity. It is possible, then, that instead of implementing an adaptation mechanism able to dynamically change the filter shape on an individual cell basis in response to stimulus skew, the neural population is structurally adapted to the overall luminance distribution of natural scenes, by properly partitioning the population between ON- and OFF-like cells \citep{ratliff10}.

Another explanation for invariant coding in response to higher-order statistics could be that the ability to adapt is limited by the ability to detect a change in the underlying statistical structure of the stimulus. For example, to detect reliably  a change in kurtosis, the neuron would need to receive some minimal number of independent stimulus samples, but this minimum might still be large, making such inferences hard. More fundamentally, it is not clear what are the actual statistics that the neurons are adapting to \citep{bonin08,simmons09}. While we commonly think in terms of contrast, skew, and kurtosis as the relevant statistical properties of stimuli, it is not obvious that the brain relies on these same measures in dealing with natural scenes. In particular, they may be poor choices in natural settings, as their values are sensitive to outliers and because they might vary in a dependent way in nature (c.f. \citet{mante05}). Perhaps then, the retina (and other neural systems) may be using  other estimators for contrast-, skew-, and kurtosis-like statistics? 

Some of these important issues could be addressed by extending our analysis either by using temporally correlated stimuli, or by heavy-tailed (or naturalistic) luminance histograms --  if they can be reproduced in the lab using display hardware with a larger dynamic range \citep{rieke09}. Both of these extensions would affect the central limit theorem argument above: in the case of temporally correlated stimulus, the samples would no longer be independently drawn and thus might not (quickly) converge to a Gaussian; in the second case, the linear filter will not be able to ``erase'' the signatures of higher-order statistics, should the decay of the luminance distribution be too shallow. As both extensions would bring the stimuli closer to the true naturalistic ones, they could provide us with a more complete  window into the nature of retinal adaptation to natural scenes.

\begin{acknowledgements}
We thank Vijay Balasubramanian, Kristina Simmons, and Jason Prentice for stimulating discussions. GT wishes to thank the faculty and students of the ``Methods in Computational Neuroscience'' course at Marine Biological Laboratory, Woods Hole. This work was supported by The Israel Science Foundation and The Human Frontiers Science Program.
\end{acknowledgements}

\begin{thebibliography}{}
%
\bibitem[Abbott et al.(1997)]{abbott97}
Abbott LF, Varela JA, Sen K \& Nelson SB (1997) Synaptic depression and cortical gain control. \emph{Science} {\bf 275:} 220--224.
%
\bibitem[Adrian(1928)]{adrian28} The Basis of Sensation. W. W. Norton, New York.
%
\bibitem[Ag\"uera y Arcas \& Fairhall(2003)]{baa1}
Ag\"uera y Arcas B \& Fairhall AL (2003) What causes a neuron to spike? \emph{Neural Comput} {\bf 15:} 1789--1807.
%
\bibitem[Atick(1992)]{atick92}
Atick JJ (1992) Could information theory provide an ecological theory of sensory processing? \emph{Network} {\bf 3:} 213--251.
%
\bibitem[Atick \& Redlich(1990)]{atick90}
Atick JJ \& Redlich AN (1990) Towards a theory of early visual processing. \emph{Neural Comput} {\bf 2:} 308--320.
%
\bibitem[Attneave(1954)]{attneave54}
Attneave F (1954) Some informational aspects of visual perception. \emph{Psychol Rev} {\bf 61:} 183--93.
%
\bibitem[Balasubramanian \& Sterling(2009)]{vijay09}
Balasubramanian V \& Sterling P (2009) Receptive fields and functional architecture in the retina. \emph{J Neurophysiol} {\bf 587:} 2753--2767.
%
\bibitem[Barlow(1961)]{barlow61}
Barlow HB (1961) Possible principles underlying the transformation of sensory messages. Sensory Communication, pp. 217--234.
%
\bibitem[Beaudoin et al.(2007)]{beaudoin07}
Beaudoin DL, Borghuis BG \& Demb JB (2007) Cellular basis for contrast gain control over the receptive field center of mammalian retinal ganglion cells. {\emph J Neurosci} {\bf 27:} 2636--2645.
%
\bibitem[Bialek \& de Ruyter van Steveninck(2005)]{rdr}
Bialek W \& de Ruyter van Steveninck RR (2005) Features and dimensions: Motion estimation in fly vision. \emph{arXiv.org:}q-bio/0505003.
%
\bibitem[Bonin et al.(2008)]{bonin08}
Bonin V, Mante V \& Carandini M (2006) The statistical computation underlying contrast gain control. \emph{J Neurosci} {\bf 26:} 6346--53.
%
\bibitem[Borst et al.(2005)]{borst05}
Borst A, Flanagin VL \& Sompolinsky H (2005) Adaptation without parameter change: dynamic gain control in motion detection. \emph{Proc Nat'l Acad Sci USA} {\bf 102:} 6172--6176.
%
\bibitem[Brenner et al.(2000)]{brenner00}
Brenner N, Bialek W \& de Ruyter van Steveninck RR (2000) Adaptive rescaling maximizes information transmission. \emph{Neuron} {\bf 26:} 695--702.
%
\bibitem[Chubb et al.(2004)]{chubb04}
Chubb C, Landy MS \& Econopouly J (2004) A visual mechanism tuned to black. \emph{Vision Res} {\bf 44:} 3223--3232.
%
\bibitem[Chander \& Chichilnisky(2001)]{chander01}
Chander D \& Chichilnisky EJ (2001) Adaptation to temporal contrast in primate and salamander retina. \emph{J Neurosci} {\bf 21:} 9904--9916.
%
\bibitem[Cover \& Thomas(1991)]{shannon}
Cover TM \& Thomas JA (1991) \emph{Elements of Information Theory}. Wiley, New York.
%
\bibitem[Fairhall et al.(2001)]{fairhall01}
Fairhall AL, Lewen GD, Bialek W \& de Ruyter van Steveninck RR (2001) Efficiency and ambiguity in an adaptive neural code. \emph{Nature} {\bf 412:} 787--792.
%
\bibitem[Fairhall et al.(2006)]{fairhall06}
Fairhall AL, Burlingame CA, Narasimhan R, Harris RA, Puchalla JL \& Berry MJ 2nd (2006) Selectivity for multiple stimulus features in retinal ganglion cells. \emph{J Neurophysiol} {\bf 96:} 2724--2738.
%
\bibitem[Geisler(2008)]{geisler08}
Geisler WS (2008) Visual perception and the statistical properties of natural scenes. \emph{Annu Rev Psych} {\bf 59:} 167--192.
%
\bibitem[Hosoya et al.(2005)]{hosoya05}
Hosoya T, Baccus SA \& Meister M (2005) Dynamic predictive coding by the retina. \emph{Nature} {\bf 436:} 71--77.
%
\bibitem[Keat et al.(2001)]{keat01}
Keat J, Reinagel P, Reid RC \& Meister M (2001) Predicting every spike: a model for the responses of visual neurons. \emph{Neuron} {\bf 30:} 803--17.
%
\bibitem[Laughlin(1981)]{laughlin81}
Laughlin S (1981) A simple coding procedure enhances a neuron's information capacity. \emph{Z Naturforsch} {\bf 36:} 910--912.
%
\bibitem[Mante et al.(2005)]{mante05}
Mante V, Frazor RA, Bonin V, Geisler WS \& Carandini M (2005) Independence of luminance and contrast in natural scenes and in the early visual system. \emph{Nature Neurosci} {\bf 8:} 1690--1697.
%
\bibitem[Meister et al.(1994)]{meister94}
Meister M, Pine J \& Baylor DA (1994) Multi-neuronal signals from the retina: acquisition and analysis. \emph{J Neurosci Methods} {\bf 51:} 95--106.
%
\bibitem[M\"uller et al.(1999)]{muller99}
M\"uller JR, Mehta AB, Krauskopf J \& Lennie P (1999) Rapid adaptation in visual cortex to the struture of images. \emph{Science} {\bf 285:} 1405--1408.
%
\bibitem[Olveczky et al.(2007)]{olveczky07}
Olveczky BP, Baccus SA \& Meister M (2007) Retinal adaptation to object motion. \emph{Neuron} {\bf 56:} 689--700.
%
\bibitem[Partridge \& Stevens(1976)]{partridge76}
Partridge LD \& Stevens CF (1976) A mechanism for spike frequency adaptation. \emph{J Physiol} {\bf 256:} 315--332.
%
\bibitem[Pillow et al.(2008)]{pillow08}
Pillow JW, Shlens J, Paninski L, Sher A, Litke AM, Chichilnisky EJ \& Simoncelli EP (2008) Spatio-temporal correlations and visual signaling in a complete neuronal population. \emph{Nature} {\bf 454:} 995--999.
%
\bibitem[Portilla \& Simoncelli(2000)]{portilla00}
Portilla J \& Simonceli EP (2000) A parametric texture model based on joint statistics of complex wavelet coefficients. \emph{Int'l J of Computer Vis} {\bf 40:} 49--71.
%
\bibitem[Ratliff et al.(2010)]{ratliff10}
Ratliff CP, Borghuis BG, Kao YH, Sterling P \& Balasubramanian V (2010) Retina is structured to process an excess of darkness in natural scenes. \emph{Proc Nat'l Acad Sci USA} {\bf 107:} 17368--73.
%
\bibitem[Rieke \& Rudd(2009)]{rieke09}
Rieke F \& Rudd ME (2009) The challenges natural images pose for visual adaptation. \emph{Neuron} {\bf 64:} 605--616.
%
\bibitem[Sadeghi(2009)]{sadeghi09}
Sadeghi KS (2009) Progress on deciphering the retinal code. Thesis, Princeton University.
%
\bibitem[Schwartz \& Berry(2008)]{schwartz08}
Schwartz G \& Berry MJ 2nd (2008) Sophisticated temporal pattern recognition in retinal ganglion cells. \emph{J Neurophysiol} {\bf 99:} 1787--98.
%
\bibitem[Shapley \& Enroth-Cugell(1984)]{shapley84}
Shapley R \& Enroth-Cugell C (1984) Visual adaptation and retinal gain controls. \emph{Progr Ret Res} {\bf 3:} 263--364.
%
\bibitem[Shapley \& Victor(1979)]{shapley79}
Shapley R \& Victor JD (1979) The contrast gain control of the cat retina. \emph{Vision Res} {\bf 19:} 431--434.
%
\bibitem[Sharpee et al.(2004)]{sharpee04}
Sharpee T, Rust NC \& Bialek W (2004) Analyzing neural responses to natural signals using maximally informative dimensions. \emph{Neural Comput} {\bf 16:} 223--250.
%
\bibitem[Simmons et al.(2009)]{simmons09}
Simmons K, Tka\v{c}ik G, Prentice JS \& Balasubramanian V (2009) What is the ``contrast'' in contrast adaptation? \emph{Front Syst Neurosci} Cosyne 2009 abstract, doi:10.3389/conf.neuro.06.2009.03.144.
%
\bibitem[Simoncelli \& Olshausen(2001)]{simoncelli01}
Simoncelli EP \& Olshausen BA (2001) Natural image statistics and neural representation. \emph{Annu Rev Neurosci} {\bf 24:} 1193--216.
%
\bibitem[Smirnakis et al.(1997)]{smirnakis97}
Smirnakis SM, Berry MJ 2nd, Warland DK, Bialek W \& Meister M (1997) Adaptation of retinal processing to image contrast and spatial scale. \emph{Nature} {\bf 386:} 69--73.
%
\bibitem[Strong et al.(1998)]{strong98}
Strong SP, Koberle R, de Ruyter van Steveninck RR \& Bialek W (1998) Entropy and information in neural spike trains. \emph{Phys Rev Lett} {\bf 80:} 197--200.
%
\bibitem[Srinivasan et al(1982)]{srinivasan82}
Srinivasan MV, Laughlin SB, Dubs A (1982) Predictive coding: a fresh view of inhibition in the retina. \emph{Proc Royal Soc B (Lond)} {\bf 216:} 427--459.
%
\bibitem[Ulanovsky et al.(2004)]{ulanovsky04}
Ulanovsky N, Las L, Farkas D \& Nelken I (2004) Multiple time scales of adaptation in auditory cortex neurons. \emph{J Neurosci} {\bf 24:} 10440--10453.
%
\bibitem[Victor(1986)]{victor86}
Victor JD (1986) The dynamics of the cat retinal X cell centre. \emph{J Physiol} {\bf 386:} 219--246.
%
\bibitem[Tka\v{c}ik et al.(2010)]{tkacik10}
Tka\v{c}ik G, Prentice JS, Victor JD \& Balasubramanian V (2010) Local statistics in natural scenes predict the saliency of synthetic textures. \emph{Proc Nat'l Acad Sci USA} {\bf 107:} 18149--54.
%
\bibitem[Tka\v{c}ik et al.(2011)]{tkacik11}
Tka\v{c}ik G, Garrigan P, Ratliff C, Mil\v{c}inski G, Klein JM, Seyfarth LH, Sterling P, Brainard DH \& Balasubramanian V (2011) Natural images from the birthplace of the human eye. \emph{PLoS One} {\bf 6:} e20409.
%
\bibitem[Tsodyks \& Markram(1997)]{tsodyks97}
Tsodyks MV \& Markram H (1997) The neural code between neocortical pyramidal neurons depends on neurotransmitter release probability. \emph{Proc Nat'l Acad Sci USA} {\bf 94:} 719--723.
%
\bibitem[Wallach et al.(2008)]{wallach08}
Wallach A, Eytan D, Marom S \& Meir R (2008) Selective adaptation in networks of heterogenous populations: model, simulation, and experiment. \emph{PLoS Comput Biol} {\bf 4:} e29.
%
\bibitem[Wark et al.(2007)]{wark07}
Wark B, Lundstrom BN \& Fairhall AL (2007) Sensory adaptation. \emph{Curr Opin Neurobiol} {\bf 17:} 423--9.
%
\end{thebibliography}
\end{document}